\documentclass[%
letterpaper,
reprint,
preprintnumbers,
nofootinbib,
floatfix,
amsmath,amssymb,
prstab,
longbibliography
]{revtex4-1}

\usepackage{dcolumn} 
\usepackage{etoolbox}
\usepackage{siunitx}
\usepackage{booktabs}
\usepackage{enumerate}
\usepackage{standalone}
\usepackage{mathtools}
\usepackage{bm}
\usepackage{microtype}

\usepackage[capitalise]{cleveref}

\usepackage{tikz}
\usetikzlibrary{arrows.meta}
\usetikzlibrary{positioning}
\usetikzlibrary{calc}
\usetikzlibrary{shapes}
\usetikzlibrary{angles,shapes.geometric}
\usetikzlibrary{pgfplots.groupplots}
\usetikzlibrary{shapes.symbols}
\usetikzlibrary{quotes}
\usetikzlibrary{chains}
\usetikzlibrary{patterns}

\AtBeginDocument{
	\heavyrulewidth=.08em
	\lightrulewidth=.05em
	\cmidrulewidth=.03em
	\belowrulesep=.65ex
	\belowbottomsep=0pt
	\aboverulesep=.4ex
	\abovetopsep=0pt
	\cmidrulesep=\doublerulesep
	\cmidrulekern=.5em
	\defaultaddspace=.5em
}

\usepackage{pgfplots} 
\pgfplotsset{compat=1.13}

\tikzset{
	block/.style={draw, very thick,	minimum height=1cm,	minimum width=1cm},		
	->/.style={thick, -{Triangle[length=2.8mm,width=1.2mm]}},
	<-/.style={thick, {Triangle[length=2.8mm,width=1.2mm]}-},
	<->/.style={thick,{Triangle[length=2.8mm,width=1.2mm]}-{Triangle[length=2.8mm,width=1.2mm]}},
}

\usepackage{amsthm}
\newtheoremstyle{remarkstyle} 
{}                    
{}                    
{}                   
{}                           
{\itshape}                   
{}                          
{.5em}                       
{\thmname{#1} \thmnumber{#2}\thmnote{, #3}: }  

\theoremstyle{remarkstyle}
\newtheorem{remark}{Remark}

\usepackage{commands}
\usepackage{pgfplotstyle}
\usepackage{phasor_plot_commands}

\newcommand{\phibo}{\phi_\text{b0}}
\newcommand{\phibolin}{\phi_\text{b0,lin}}
\newcommand{\phibocirc}{\phi_\text{b0,circ}}

\begin{document}
	
\newcommand{\ESSRFQ}{ESS RFQ}
\newcommand{\ESSDTL}{ESS DTL}
\newcommand{\ESSMedBeta}{ESS Medium-$\beta$}
\newcommand{\ESSHighBeta}{ESS High-$\beta$}
\newcommand{\XFELTesla}{X-FEL (TESLA)}
\newcommand{\LCLSIITesla}{LCLS II (TESLA)}	
\newcommand{\CESR}{CESR (Phase III)}	

\newcommand{\imgpath}{.}

\preprint{}

\title{Energy-based parameterization of accelerating-mode dynamics}

\author{Olof Troeng}
\thanks{E-mail: \texttt{oloft@control.lth.se}}%

\affiliation{%
	Department of Automatic Control, Lund University, Sweden\\ 
}

\date{\today} 

\renewcommand{\RF}{\text{rf}}
\renewcommand{\DC}{\text{dc}}

\renewcommand{\bz}{\pmb{z}}

\begin{abstract}
We propose a parameterization of the accelerating-mode dynamics of accelerating cavities in terms of energy, power, and current. This parameterization avoids many confusing features of the popular equivalent-circuit-based parameterization where a fictitious generator current is introduced.
To further simplify analysis and understanding of the accelerating-mode dynamics we also propose a convenient normalization and use phasor diagrams for illustrations.
\end{abstract}

\pacs{Valid PACS appear here}
\maketitle

\section{Introduction}
Radio-frequency  (\rf{}) particle accelerators accelerate bunches of charged particles using oscillating electromagnetic fields that are confined in \rf{} cavities. 
Analysis of field transients and field control algorithms require a model of the field dynamics.
\Rf{} cavities have infinitely many electromagnetic eigenmodes, but in many situations it is sufficient to model the so-called accelerating mode intended for particle acceleration.

Since the accelerating mode provides a voltage to the charged particles, and the particle beam corresponds to a current, it is natural to model the accelerating mode as an equivalent (electric) circuit.
This is the standard approach to model accelerator cavities \cite{Schilcher1998,Tuckmantel2011}%
\footnote{Much of the existing literature has an emphasis on steady-state relations \cite{Padamsee2008,Wangler2008}, which are not helpful for field control analysis.}
and is also common in \rf{} engineering \cite{Montgomery1948,Pozar2009}.

However, equivalent-circuit-based parameterizations of the accelerating mode have a number of inconvenient features\footnote{\label{fn:tuckmantel}
For example, as was noted by T\"{u}ckmantel \cite{Tuckmantel2011}, ``for considerations where $Q_\ext$ varies---as for a variable coupler e.g.
in context with \rf{} vector feedback loop gain---or where $(R/Q)$ varies---as when
particles of different speed $v = \beta c$ pass the same cavity /.../ the model currents cannot be considered constant; they have to be re-adapted each
time $Q_\ext$ or $(R/Q)$ change''.} \cite{Tuckmantel2011}.
These mainly stem from that the \rf{} drive needs to be considered as a virtual current to fit it into the equivalent-circuit framework. 
A parameterization that avoids these issues was proposed by Haus for modeling optical cavities \cite{Haus1983}.
In Haus' parameterization the amplitudes of the accelerating mode and the \rf{} drive are taken as the square root of the mode energy and the square root of the forward power, respectively.
These quantities are, unlike those in equivalent-circuit-based parameterizations, defined independently of the beam velocity and the strength of the cavity--waveguide coupling.

In this paper we extend Haus' energy-based parameterization \cite{Haus1983} to include beam loading, which enables it to model accelerating cavities.
We then discuss its advantages over equivalent-circuit-based parameterizations.
We also propose a normalization of the cavity dynamics that simplifies the analysis of field control algorithms.
Throughout, we illustrate the discussed concepts with a somewhat novel type of phasor diagrams.

\begin{figure}
	\centering
	\begin{tikzpicture}
\coordinate (cavpos) at (0, 0);

\node[draw, cylinder, ultra thick, minimum width=1.1cm, minimum height=2.8cm] at (cavpos) (cav) {};

\draw[-, blue, thick, decorate,decoration={snake,amplitude=4mm,segment length=4.4mm}] ($(cav.west)+ (0.3,0)$) -- ($(cav.east)-(0.4,0)$);

\node[align=left, anchor=north] at ($(cavpos.south) + (0,-0.8)$) {Envelope of the accelerating\\cavity mode: $\color{blue}\bA$ {\color{gray} ($\bV$)}};

\coordinate (cavport-east) at ($(cav.north east) + (0.45, 0)$);

\draw[very thick, double distance=5mm, align=left] (cavport-east)+(0,2) --
 node[left,pos=0.5,xshift=-7pt,align=left] {\Rf{} drive: ${\color{ig-color}\Fg}$  {(\color{gray}$\Ig$)}\\\footnotesize(Forward wave) }
 node[right,pos=0.28,xshift=6pt] {\footnotesize Reverse wave: $\mathbf{R_g}$}
(cavport-east)  ;

\draw[->, very thick, decorate,decoration={snake,amplitude=0.8mm,post length=2.5mm}, ig-color] ($(cavport-east) + (0,1.7)$) -- +(0,-1.5);

\draw[->, decorate,decoration={snake,amplitude=0.3mm,post length=2.5mm}, black] ($(cavport-east) + (0.16,1.0)$) -- +(0,1);

\draw[thick] (cavport-east) -- +(0, 0.15)
-- +(0,-0.15);

\draw[-{Triangle[length=5mm,width=2mm]}, ib-color, ultra thick,
line width=1mm, line cap=round, dash pattern=on 0mm off 4.4mm
] ($(cav.west) - (1.15,0)$) 
-- node[below, anchor=north west, pos=1,align=left,yshift=-3pt,xshift=-40pt,color=black] {Beam current: ${\color{ib-color}\Ib}$ {\color{gray}($\IbRF$)}} ($(cav.east) + (1.5,0)$);

\end{tikzpicture}
	\caption{Illustration of an accelerating \rf{} cavity coupled to a a waveguide. The cavity field is established and maintained by the forward wave $\Fg$ that is provided by an \rf{} amplifier (\emph{g}enerator).
		The clearly colored letters denote the complex envelopes that are used in the proposed energy-based parameterization.
		Parenthesized gray letters indicate the variables that are typically used in  equivalent-circuit-based parameterizations.}
	\label{fig:cavity_waveguide}
\end{figure}
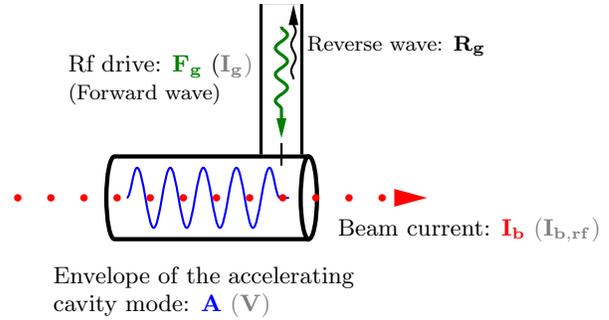

\vspace{1em}
\noindent\emph{Notation and assumptions:} 
(1) The considered system is shown in \cref{fig:cavity_waveguide}. We restrict our attention to the accelerating cavity mode and will not consider parasitic modes.
(2) \Rf{} signals are represented by their complex envelopes (equivalent baseband signals), which are denoted by bold letters.
(3) Particle bunches are assumed to be point-like (it is straightforward to include a relative bunch form factor \cite{Tuckmantel2011}).

\section{Background: Equivalent-Circuit Based Parameterizations}

Two popular, equivalent-circuit-based parameterizations in the existing literature are: Tückmantel's \cite{Tuckmantel2011}\footnote{Eq. (53), with minor modifications for consistency.},
\begin{subequations}
	\begin{equation}
	\d{\bV}{t} = - \! \left[ \frac{\omega_a }{2Q_\ext} \!+\! \frac{\omega_a }{2Q_0} \!-\! i \Dw \right] \bV \!+\! \omega_a \frac{r_\circ}{Q} \Ig + \omega_a \frac{1}{2} \frac{r_\circ}{Q} \IbRF,
	\label{eq:cav_eq_tuckmantel}
	\end{equation}
	and Schilcher's \cite{Schilcher1998}\footnote{Eq. (3.49) interpreted for complex signals, with $\textbf{I} = 2\Ig + \IbRF$.},
	\begin{equation}
	\d{\bV}{t}  = (-\wbw + i\Delta \omega)\bV + R_L\wbw \left(2\bIg + \IbRF \right).
	\label{eq:cav_eq_schilcher}
	\end{equation}
	
	In the above equations, $\bV$ denotes the complex envelope of the effective accelerating voltage of the cavity field; $\Ig$ denotes the ``generator current'' which models the \rf{} amplifier drive; and $\IbRF$ denotes the \rf{} component of the beam current ($\abs{\IbRF} = 2I_\DC$ \cite[A4]{Schilcher1998}), see Tables \ref{tab:cavity_parameter_comparison}a and \ref{tab:cavity_parameter_comparison}b for a complete list of the quantities involved.
	The normalized shunt impedance $r_\circ/Q$ in (\ref{eq:cav_eq_tuckmantel})  is defined with respect to the \emph{equivalent-circuit convention}, and so is the loaded resistance $R_L$ in (\ref{eq:cav_eq_schilcher}). There is also the \emph{linac convention} for which $r/Q =  2(r_\circ/Q)$,  \cite{Tuckmantel2011}.
	
	Note that the generator current $\Ig$ is a fictitious quantity that is introduced to make the \rf{} drive term fit into the equivalent-circuit framework.
	An additional relation is needed for how $\Ig$ relates to the drive power $\Pg$, which is the physical quantity of interest,
	\begin{equation}
	\Pg = \frac{1}{2} \frac{r_\circ}{Q} Q_\ext \abs{\bIg}^2.
	\label{eq:generator_power}
	\end{equation}
\end{subequations}

\section{Energy-Based Parameterization}
\label{sec:proposed_parameterization}
Let the state of the accelerating cavity mode be quantified by the complex-valued mode amplitude $\bA$, with $\abs{\bA}^2$ equal to the stored mode energy ($\bA$ has units \si{\sqrt{J}}).
The mode amplitude $\bA$ is related to the effective accelerating voltage via
\[
\bV = \alpha \bA,
\]
where $\alpha =  \sqrt{\omega_a (r/Q)}$ quantifies the coupling between the cavity field and the beam.
Recall that $\alpha$, just like $(r/Q)$, depends on the beam velocity \cite{Wangler2008}.
With these definitions the dynamics of the accelerating mode can be written
\begin{equation}
\d{\bA}{t}  = (-\gamma + i\Dw) \bA  + \sqrt{2\gamma_\ext} \Fg + \frac{\alpha}{2}  \Ib,
\label{eq:cav_eq_proposed}
\end{equation}
where $\gamma = \gamma_0 + \gamma_\ext$ is the decay rate of the cavity field ($\gamma_0$ corresponds to resistive losses and $\gamma_\ext$  to decay through the power coupler); $\Fg$ is the envelope of the forward wave from the \rf{} amplifier with $\abs{\Fg}^2$ equal to the power in the wave ($\Fg$ has units $\si{\sqrt{W}}$); and $\Ib$ is the \mbox{beam-loading phasor}, with $\abs{\Ib}$ equal to the dc beam current.

\begin{remark}
If an equation in $\bV$ is desired, multiplying (\ref{eq:cav_eq_proposed}) by $\alpha$ gives
\begin{equation}
\d{\bV}{t}  = (-\gamma + i\Dw) \bV  + \alpha\sqrt{2\gamma_\ext} \Fg + \frac{\alpha^2}{2}  \Ib.
\label{eq:cav_eq_proposed_voltage}
\end{equation}
Most of the advantages mentioned in the next subsection apply to this parameterization as well.
\end{remark}

\begin{remark} 
	In the proposed parameterization (\ref{eq:cav_eq_proposed}) we considered the factor $\gamma$ as a decay rate rather than as a bandwidth as in (\ref{eq:cav_eq_schilcher}).
	This is consistent with the laser literature \cite{Siegman1986} and makes it natural to write the total decay rate as $\gamma = \gamma_0 + \gamma_\ext$;  a relation that would be less intuitive in terms of bandwidths.
	For frequency-domain considerations one should, of course, think of $\gamma$ as a bandwidth.
\end{remark}

\begin{remark} 
The quality factor (Q factor) of an oscillator quantifies its decay in terms of oscillation periods. Q factors are popular  \cite{Merminga1996,Padamsee2008,Tuckmantel2011,Schilcher2007} for quantifying the decay of rf cavity modes, e.g., $Q_0$, $Q_\ext$, and $Q_L$ in Table~\ref{tab:cavity_parameter_comparison}b. 
However, in most situations where the field dynamics are of interest, such as field control, it is the absolute timescales that are of interest. In these situations Q factors provide little information on their own (before division with $\wRF$). The decay rates $\gamma_0$, $\gamma_\ext$, and $\gamma$ in Table~\ref{tab:cavity_parameter_comparison}c, on the other hand, capture all relevant information and are meaningful in their own right.
\end{remark}

\begin{remark} 
As indicated in \cref{fig:cavity_waveguide}, there is a reverse wave present in the waveguide.
The complex envelope of the reverse wave (units \si{\sqrt{W}}) is given by (see Appendix \ref{sec:reverse_wave})
\begin{equation}
\Rg = -\Fg + \sqrt{2\gamma_\ext}\bA.
\end{equation}
Under ideal steady-state conditions $\Rg = 0$, see Sec.~\ref{sec:opt_coupling_detuning}.
\end{remark}

\begin{table}
	\centering
	\caption{Physical quantities in the equivalent-circuit-based parameterizations (\ref{eq:cav_eq_tuckmantel})/(\ref{eq:cav_eq_schilcher}) and in the proposed parameterization (\ref{eq:cav_eq_proposed}).
		The rightmost column contains the quantities expressed in the parameters of the other parameterization.}
	\vspace{0.8em}
	\begin{tabular*}{\columnwidth}{@{}cclr@{}}
		\toprule
		\multicolumn{4}{l}{\textbf{a) Quantities common to
				(\ref{eq:cav_eq_tuckmantel})/(\ref{eq:cav_eq_schilcher}) and (\ref{eq:cav_eq_proposed})}} \\[0.2em]
		\midrule\\[-0.9em]
		$\omega_a$ & \si{\radian/\second} & \multicolumn{2}{l}{Resonance frequency of the accelerating mode} \\
		$\Delta \omega$ & \si{\radian/\second} &
		\multicolumn{2}{l}{Detuning of the accelerating mode, $= \omega_a - \omega_\RF$} \\
		
		\toprule
		\multicolumn{4}{l}{\textbf{b) Quantities in
				(\ref{eq:cav_eq_tuckmantel})/(\ref{eq:cav_eq_schilcher}) }} \\[0.2em]\midrule\\[-0.9em]
		$\bV$ & \si{V} & Accelerating voltage & $\bV = \alpha \bA$ \\
		$\mathbf{I_g}$ & \si{A} & Generator current & $2\sqrt{2\gamma_\ext}/\alpha \Fg$ \\ 
		$\IbRF$ & \si{A}  & Beam current (RF component) & $2\Ib$  \\
		\midrule
		$Q_0$ & -- & Unloaded quality factor & $\omega_a / (2 \gamma_0)$ \\
		$Q_\ext$ & -- & External quality factor & $\omega_a / (2 \gamma_\ext)$ \\
		$\beta$ & -- &  Coupling factor, $ = Q_0 / Q_\ext$ & $\gamma_\ext / \gamma_0$ \\
		$Q_L$ & -- & \multicolumn{2}{l}{Loaded quality factor, $= Q_0 / (\beta + 1)$ \hfill $\omega_a / (2 \gamma)$\hspace*{-0.1cm}} \\
		$\wbw$ & \si{\radian/\second} & Half bandwidth, $= \omega_a/(2Q_L)$ & $\gamma$ \\
		$r/Q$ & \si{\ohm} & 		Normalized shunt impedance, &  $\alpha^2/\omega_a$ \\
		& & \,\, linac convention & \\
		$r_\circ/Q$ & \si{\ohm} &
		Normalized shunt impedance, & $\alpha^2/(2\omega_a)$ \\
		& & \multicolumn{2}{l}{\,\, equiv.-circuit convention, $=\!(r/Q)/2$} \\
		$R_L$ & \si{\ohm} & 	Loaded shunt impedance, & $\alpha^2/(4\gamma)$ \\
		& & \qquad $= (r_\circ/Q)\cdot Q_0 /(1+\beta)$\\		
		\toprule
		\multicolumn{4}{l}{\textbf{c) Quantities in (\ref{eq:cav_eq_proposed})}} \\[0.2em]
		\midrule
		$\bA$ & \si{\sqrt{J}}  & Mode amplitude & $\bV / \sqrt{\omega_a (r/Q)}$ \\
		$\Fg$ & \si{\sqrt{W}} & \multicolumn{2}{l}{\Rf{} drive \hfill $\sqrt{(r/Q) Q_\ext / 4} \bIg$\!\!} \\
		$\Ib$ & \si{A} & Beam current & $\IbRF / 2$ \\[0.1em]
		\midrule
		$\gamma_0$ & \si{1/s} & Resistive decay rate  & $\omega_a / (2 Q_0)$ \\
		$\gamma_\ext$ & \si{1/s} & External decay rate & $\omega_a / (2 Q_\ext)$  \\
		$\gamma$  & \si{1/s} &Total decay rate, $= \gamma_0 + \gamma_\ext$ & $\wbw$ \\
		& & \quad (half bandwidth) &  \\
		$\alpha$ & \si{V\!/\!\sqrt{J}} & Field--beam coupling parameter  & $\sqrt{\omega_a (r/Q)}$ \\		
		\bottomrule
	\end{tabular*}
	\label{tab:cavity_parameter_comparison}
\end{table}

\subsection*{\mbox{Comparison of the energy-based parameterization (\ref{eq:cav_eq_proposed})}\\to equivalent-circuit-based parameterizations}
\label{sec:model_comparison}

With Table~\ref{tab:cavity_parameter_comparison} it is easy to verify that the three parameterizations (\ref{eq:cav_eq_tuckmantel}), (\ref{eq:cav_eq_schilcher}) and (\ref{eq:cav_eq_proposed}) are equivalent.
Note that also field control requirements of the form $x\%$ amplitude error and $y^\circ$ phase error are identical for $\bV$ and $\bA$

Below we give some pros and cons of the proposed parameterization  	(\ref{eq:cav_eq_proposed}) relative to the parameterizations (\ref{eq:cav_eq_tuckmantel}) and (\ref{eq:cav_eq_schilcher}).\\

\noindent\emph{Advantages} of the proposed parameterization (\ref{eq:cav_eq_proposed}) are:

\begin{enumerate}[1.]
	\item The dynamic equation is cleaner. Compare for example the expression in (\ref{eq:generator_power}) to $P_\text{g} = \abs{\Fg}^2$.
	
	\noindent\emph{Remark:}
	Quantifying signal amplitudes in terms of square root of power, as for $\Fg$, is common in \rf{} engineering. It is with respect to such power waves that scattering matrices are defined \cite{Pozar2009}.
	
	\noindent
	\emph{Example:} The \rf{} drive power necessary to maintain an accelerating  voltage $\Vo = \alpha\bA_0$, while accelerating a beam modeled by $\Ibo$, is easily found from (\ref{eq:cav_eq_proposed}) as
	\[
	P_\text{g} = \frac{1}{2\gamma_\ext} \abs{
		(-\gamma + i\Dw) \Vo / \alpha  + \frac{\alpha}{2} \Ibo
	}^2.
	\]
	This expression is more convenient and easier to remember than (\ref{eq:cav_eq_tuckmantel}/\ref{eq:cav_eq_schilcher}) together with (\ref{eq:generator_power}).
		
	\item The impact of changes to the cavity--waveguide coupling $\gamma_\ext$ and cavity--beam coupling $\alpha$ are transparent. The same cannot be said for the parameterizations (\ref{eq:cav_eq_tuckmantel})/(\ref{eq:cav_eq_schilcher}).
	
	\noindent\emph{Example:} The quantity $\Fg$ in (\ref{eq:cav_eq_proposed}) that represents the \rf{} drive is independent of $\gamma_\ext$. Contrast this to the definition of $\Ig$, implicitly given by \eqref{eq:generator_power}, that includes both $Q_\ext$ and $r/Q$.
	Thus, the parameterization (\ref{eq:cav_eq_proposed}) avoids the issue
	mentioned in footnote \ref{fn:tuckmantel}.

	\noindent
	\emph{Example:} From (\ref{eq:cav_eq_schilcher}) we see that the transfer function from beam-current variations to cavity field errors is
	\[
	G_{\Ib \rightarrow \bm{e}}(s) = \frac{\wbw}{s + \wbw} R_L.
	\]
	Recognizing the first factor as a low-pass filter with bandwidth $\wbw$, one is led to believe that reducing $\wbw$ reduces the field errors from beam-current ripple.
	This is incorrect, however, since $R_L$ depends inversely on $\wbw$.
	This confusion does not arise from (\ref{eq:cav_eq_proposed}).
		
	\item The mode amplitude $\bA$ depends only the cavity field, while the effective cavity voltage $\bV$ also depends on the beam velocity \cite{Wangler2008}. Without a given beam velocity, the effective accelerating voltage $\bV$ is not well-defined.
	
	For electron linacs, one could assume that the beam velocity equals $c$ and hence uniquely quantify the amplitude of the accelerating mode by its effective voltage.
	But the amplitudes of parasitic same-order modes of multi-cell cavities can obviously not be quantified this way.
	For example, the same-order modes of the TESLA cavity all have a negligible coupling to the beam, i.e., zero effective voltage, but they are crucial to consider in field-control analysis.
	
	
	\item The equation \eqref{eq:cav_eq_proposed} can be derived using basic properties of  Maxwell's equations (see the Appendix). This arguably allows for a better understanding of how the model parameters relate to physical cavity properties.
	
	
\end{enumerate}

\noindent\emph{Disadvantages} of the proposed parameterization \eqref{eq:cav_eq_proposed} are:

\begin{enumerate}[1.]
\item The parameterization \eqref{eq:cav_eq_proposed} does not explicitly contain the accelerating voltage $\bV$, which is arguably the most important quantity from a beam perspective.
If it is necessary with an equation in $\bV$ one may use \eqref{eq:cav_eq_proposed_voltage} which retains many of the advantages of \eqref{eq:cav_eq_proposed}.

Overall, there is less reason to use the energy-based parameterization \eqref{eq:cav_eq_proposed} when the focus is on beam stability, as is often the case in circular machines. It is from an \rf{} or field stability perspective that \eqref{eq:cav_eq_proposed} brings helpful intuition.

\item The parameters in \eqref{eq:cav_eq_proposed} are rarely used in cavity specifications.
The relationships between these parameters and those in  (\ref{eq:cav_eq_tuckmantel})/(\ref{eq:cav_eq_schilcher}) are, however, easily found with the help of Table~\ref{tab:cavity_parameter_comparison}.
\end{enumerate}

\subsection*{Examples of cavity parameters}
Typical parameters and operating points for some different cavities, expressed in the quantities of (\ref{eq:cav_eq_proposed}), are given in Table~\ref{tab:cav_params_physical}.

\begin{table}
	\setlength{\tabcolsep}{2pt}
	\caption{Parameters for some different cavities in high-energy linacs \cite{XFELTDR2007,Doolittle2016} and an electron storage ring \cite[Sec.~17.7]{Padamsee2008}. The first group of parameters are inherent to the cavity (although $\gamma_\ext$ could be tunable), the second group of parameters gives the nominal operating point, and the third group are derived quantities of interest. We have approximated $\gamma_0$ with $0$ for superconducting cavities, which is reasonable from an \rf{} system perspective.}
	\small
	\resizebox{\columnwidth}{!}{
	\begin{tabular}{@{}l@{\hskip -2pt}S@{\hskip 4pt}S@{\hspace{4pt}}c|SSc|Sc@{}}
		\toprule
		& {$\gamma_0/2\pi$} & {$\gamma_\text{ext}/2\pi$}   & $\alpha$  & {$A_0$} & {$I_\text{DC}$} & $\phibolin$ & $\Delta {E}$ & {$\abs{\Fgo}^2$} \\
		\textbf{Cavity} & \si{\kilo\hertz} & {\si{\kilo\hertz}} & {\si{MV/\sqrt{J}}}  & \si{\sqrt{J}} & {\si{mA}} & \si{\degree} & \si{MeV} & \si{kW} \\
		\midrule
		\ESSRFQ & 24 & 36 &  3.1  & 1.6 & 62.5 & \num{-45}$^\text{a}$ & 3.5 & 1000 \\
		\ESSDTL & 3.2 & 8.8 &  3.7 & 5.2 & 62.5 & \num{-25} & 18 & 2200\\
		\ESSMedBeta &  0 & 0.5 &  1.3 & 11.2 & 62.5 & \num{-15} & 14 & 900\\
		\XFELTesla &  0 & 0.14 &  2.9 & 8.1 & 5.0 & $\approx\!0$ & 24 & 120\\
		\LCLSIITesla &  0 & 0.016 &  2.9 & 5.5 & 0.1 & $\approx\!0$ & 16 & 2.5\\
		\midrule
		\CESR &  0 & 1.2 &  0.5 & 5.7 & 500 & 70 & 1.0 & 500\\  		
		\bottomrule
		\\[-0.8em]
		\multicolumn{9}{r}{\parbox{1.2\columnwidth}{\raggedright\small $^\text{a}$ The beam loading in a radio-frequency quadrupole is always relative to the phase of the accelerating mode, i.e., $\Ib(t) = I_b(t) \me^{i(\pi-\phi_{b, \text{rel}})} \cdot \me^{i\angle \bA(t)}$. The lumped value $\phi_{b, \text{rel}}$ is typically not given in RFQ specifications; the value $-45^\circ$ is an estimate by the author based on the parameters of the individual RFQ cells \cite{Ponton2013}.}}
	\end{tabular}
	}
	\label{tab:cav_params_physical}
\end{table}

\section{Phasor diagrams}
\label{sec:phasor_diagram}

\begin{figure}
	\centering
	\providecommand{\anglearrowtip}{{Triangle[length=1mm,width=1mm]}}
\begin{tikzpicture}
\newcommand{\ibcoord}{-1, -0.6}%
\newcommand{\decaycoord}{-1, 0.12}%

\drawstatediagram

\drawderivativediagram	

\node[above, gray] at ($(O1) + (2.25,0)$) {\small Re};
\node[right, gray] at ($(O1) + (0,1.5)$) {\small Im};

\node at (ig-absolute) [above]
{\color{ig-color}$\sqrt{2\gamma_\ext}\Fg$};
\node at (ib-absolute) [below]
{\color{ib-color}$\displaystyle \frac{\alpha}{2} \Ib$};	
\node[] at ($(decay-absolute) + (-0.15,0.32)$) {\color{decay-color}$(-\gamma\!+\!i\Dw)\bA$};

\coordinate (origin) at (O2);
\coordinate (posxaxis) at ($(O2) + (1,0)$);				
\coordinate (negxaxis) at ($(O2) + (-1,0)$);			
\pic [-\anglearrowtip, draw, "$\phi_\text{g}$", angle eccentricity=1.3,  angle radius = 1.3cm] {angle=posxaxis--origin--ig-absolute};
	
\node[above, gray] at ($(O2) + (2.25,0)$) {\small Re};
\node[right, gray] at ($(O2) + (0,1.5)$) {\small Im};

	\end{tikzpicture}	
	\caption{Phasor diagrams for visualizing  the dynamics of the accelerating mode in equation (\ref{eq:cav_eq_proposed}).
		\emph{Left:} Phasor for the mode amplitude.
		\emph{Right:} Phasors that affect the time derivative of the mode amplitude; in this figure they sum to zero which indicates steady-state operation.}
	\label{fig:phasor_plot_example}
\end{figure}
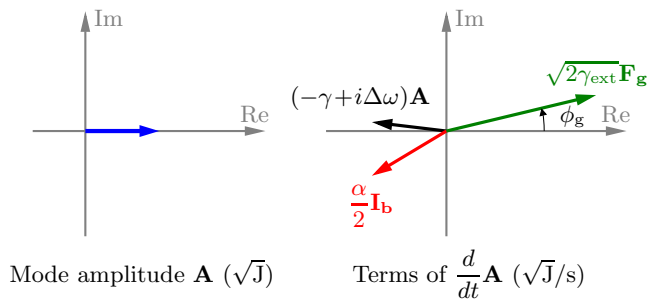

To better understand the dynamics of the accelerating mode 
it is helpful to use phasor diagrams as in \cref{fig:phasor_plot_example}.
To avoid clutter, the phasor for the cavity field (blue) is shown separately from the phasors that affect its derivative.
We will refer to them as the field-decay phasor, the rf-drive phasor (green) and the beam-loading phasor (red).

\begin{remark}
	Phasor diagrams in the previous literature typically show the phasors for the \rf{} drive and the beam loading together with their so-called induced voltages (i.e., their steady-state effects on the cavity field) in a single diagram \cite{Wangler2008,Padamsee2008,Wiedemann2007}.
	For cavity field control, one needs to understand how \emph{variations} of the phasors $\Fg$ and $\Ib$ affect the cavity field.
	In this regard, the induced voltages are of little interest.
	Leaving the them out and displaying the mode amplitude separate from the terms that affect its time derivative reduces clutter.
\end{remark}

\begin{remark} 
	\label{rmk:ref_phase}
	In previous literature, the reference phase is often chosen so that the beam-loading phasor $\Ib$ is oriented along the negative real axis \cite[p. 348]{Wangler2008}.
	This is reasonable, since after all, the beam phase is the  reference relative to which the cavity field should be controlled.
	However, from a field-control perspective (and in particular for linacs), where the objective is to keep the cavity field close to a setpoint and beam variations act as  disturbances, it is arguably more natural to choose the reference phase so that the cavity-field phasor lies on the positive real axis.
	
	With this  convention we get a nice symmetry in the phasor diagrams for optimally tuned cavities (\cref{fig:phasor_plot_sc_cavity}), with the cavity-field phasor and the rf-drive phasor lying on the real axis.
	Also, amplitude variations and (small) phase variations of these two phasors correspond to variations of their real and imaginary parts, respectively.
\end{remark}

\begin{remark}
For the particle bunches to experience acceleration and longitudinal focusing, the beam-loading phasor must lie in the second quadrant for circular machines operating above transition and in the third quadrant for linacs and circular machines operating below transition \cite{Wiedemann2007,Wangler2008,Padamsee2008}.
\end{remark}

\section{Power-optimal coupling and detuning}
\label{sec:opt_coupling_detuning}
In this section we compute the detuning $\Dw$ and the coupling $\gamma_\ext$ that minimize the rf drive power during steady-state operation.
These are standard calculations \cite{Schilcher1998,Wangler2008,Padamsee2008,Tuckmantel2011}, but we go through them to show what they look with the parameterization (\ref{eq:cav_eq_proposed}) and because we need the results in the next section.
Note that these calculations that use \eqref{eq:cav_eq_proposed} are arguably more clear than those in the previous literature.

Assume that the nominal mode amplitude is given by\footnote{Recall Remark~\ref{rmk:ref_phase}.} $\bAo = A_0 > 0$ and that the nominal beam-loading phasor is given by $\Ibo$.
The corresponding stationary \rf{} drive $\Fgo$ satisfies
\begin{equation}
0  = (-\gamma + i\Dw) A_0  + \sqrt{2\gamma_\ext} \Fgo + \frac{\alpha}{2} \Ibo,
\label{eq:baseband_cav_eq_steady_state}
\end{equation}
which gives the \rf{} drive power
\begin{align*}
\Pgo &= \abs{\Fgo}^2 \notag
= \frac{1}{2 \gamma_\ext}
\abs{
	(-\gamma + i\Dw) A_0
	+
	\frac{\alpha}{2} \Ibo
}^2
\notag.
\end{align*}
We see that the power consumption $\Pgo$ is minimized by making the imaginary part of the expression within the absolute value zero by choosing the detuning as
\begin{equation}
\Dw^\star = - \frac{\dfrac{\alpha}{2}\cdot  \Im\,\Ibo }{A_0}.
\label{eq:optimal_detuning}
\end{equation}

If the cavity is optimally tuned, then (at steady-state) the rf-drive phasor lies on the positive real axis, and the imaginary parts of the decay phasor and the beam-loading phasor have equal magnitudes but opposite signs.
It can be seen that \cref{fig:phasor_plot_sc_cavity,fig:phasor_plot_nc_cavity} correspond to optimally tuned cavities, but that \cref{fig:phasor_plot_example} does not.

With power-optimal detuning we have
\[
\Pgo \Big|_{\Dw = \Dw^\star} = \frac{1}{2 \gamma_\ext}
\left((\gamma_0 + \gamma_\ext) A_0 - \frac{\alpha}{2} \Re\, \Ibo \right)^2.
\]
Minimizing this expression with respect to $\gamma_\ext$, gives the power-optimal coupling coefficient\footnote{We wish to minimize $2f(x) = (ux+v)^2/x=u^2 x+2uv + v^2/x$ wrt $x > 0$. Differentiating gives $2f'(x) = u^2 - v^2/x^2$, from which we find the optimal point $x^\star = v/u$, at which $f(x^\star) = 2vu$.}
\begin{equation}
\gamma_\ext^\star = \gamma_0 - \frac{\dfrac{\alpha}{2} \Re\, \Ibo}{A_0}.
\label{eq:optimal_gamma_ext}
\end{equation}

Thus, given $A_0$ and $\Ibo$, the minimal power consumption equals
\[ \Pgo^\star = 2\gamma_0 A_0^2 - \alpha A_0 \cdot \Re\, \Ibo. \]
That is, all energy in the forward wave is either dissipated in the cavity walls or transferred to the particle beam---no power is wasted in the reverse wave.
The total decay rate, assuming optimal coupling, is given by
\begin{equation}
\gamma = 2\gamma_0 - \frac{\dfrac{\alpha}{2} \Re\, \Ibo}{A_0}.
\label{eq:total_decay_constant}
\end{equation}

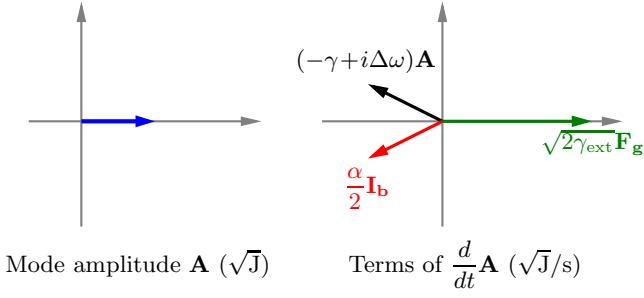
\begin{figure}
	\centering
	\newcommand{\ibcoord}{-1, -0.5}%
\newcommand{\decaycoord}{-1, 0.5}%
\begin{tikzpicture}

\drawstatediagram

\drawderivativediagram

\node at (ig-absolute) [below]
{\color{ig-color}$\sqrt{2\gamma_\ext}\Fg$};
\node at (ib-absolute) [below]
{\color{ib-color}$\dfrac{\alpha}{2} \Ib$};	
\node[] at ($(decay-absolute)+(-0.02,0.3)$) {\color{decay-color}$(-\gamma\!+\!i\Dw)\bA$};	

\end{tikzpicture}
	\caption{Phasor diagram for a superconducting cavity that is optimally tuned and optimally coupled.}
	\label{fig:phasor_plot_sc_cavity}
\end{figure}
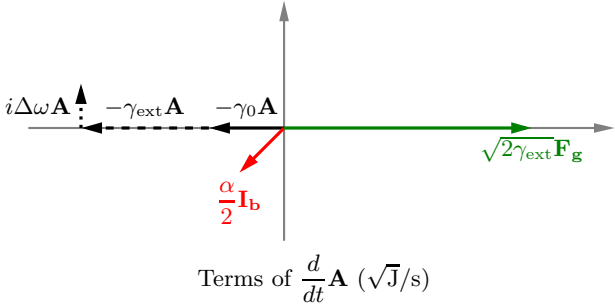
\begin{figure}
	\centering
	\newcommand{\nostateplot}{}
\newcommand{\noderivativetext}{}
\providecommand{\ibRe}{-0.6}%
\providecommand{\ibIm}{-0.6}%
\providecommand{\decayresistive}{-1.0}%
\providecommand{\decaycoord}{-2, 1}%

\begin{tikzpicture}
\coordinate (O2) at (4.8, 0);
\coordinate (dl2) at ($(O2) + (-3.4, -1.5)$); 
\coordinate (ur2) at ($(O2) + (4.2, 1.5)$); 

\draw[->, phasoraxisstyle] (O2 |- dl2) -- ($(O2 |- ur2) + (0, 0.2)$);
\draw[->, phasoraxisstyle] (dl2 |- O2) -- ($(ur2 |- O2) + (0.2, 0)$);

\coordinate (ib) at ($(O2) + (\ibRe,\ibIm)$); 
\coordinate (decay) at ($(O2) + (\decaycoord)$);
\coordinate (decay-resistive) at ($(O2) + (\decayresistive,0)$);
\coordinate (decay-ext) at ($(decay-resistive) + (\decayresistive-0.7,0)$);
\coordinate (ig) at ($2*(O2) +(-\ibRe,0) - (decay-ext)$);

\draw[phasor, decay-color] (O2) -- node[above] {$-\gamma_0\bA$} (decay-resistive);
\draw[phasor, decay-color, dashed] (decay-resistive) -- node[above] {$-\gamma_\ext\bA$} (decay-ext);
\draw[phasor, decay-color, dotted] (decay-ext) -- node[left] {$i\Dw\bA$} ++(0, -\ibIm);

\draw[phasor,ib-color] (O2) -- (ib);
\draw[phasor,ig-color]     (O2) -- (ig);

\node at (ig) [below]
{\color{ig-color}$\sqrt{2\gamma_\ext}\Fg$};
\node at (ib) [below]
{\color{ib-color}$\dfrac{\alpha}{2} \Ib$};

\node at ($(dl2)!0.5!(ur2) + (0,-2.0)$) [align=center] 
{\small Terms of $\displaystyle \d{}{t} \bA$ \sibrac{\sqrt{J}\per\second}};


\end{tikzpicture}	
	\caption{Phasor diagram with the terms of the time derivative of the cavity field.
		The considered cavity is optimally tuned ($\Dw  A_0= -(\alpha/2)\Im\,\Ibo$), optimally coupled ($\gamma_\ext A_0 = \gamma_0 A_0 - (\alpha/2)\Re\, \Ibo$), and normal conducting ($\gamma_0 > 0$); compare (\ref{eq:optimal_detuning}) and (\ref{eq:optimal_gamma_ext}).}
	\label{fig:phasor_plot_nc_cavity}
\end{figure}

\begin{remark}
We have found that it sometimes gives intuition to think of the second term in \eqref{eq:optimal_gamma_ext} as the \emph{decay rate of the cavity field due to beam loading} (at the nominal operating point). This motivates the definition
\begin{equation}
\gamma_\text{beam} =
\gamma_\text{beam}(A_0, \Ibo) \coloneqq
-\frac{\dfrac{\alpha}{2} \Re\, \Ibo}{A_0}.
\label{eq:def_gamma_beam}
\end{equation}
Using (\ref{eq:def_gamma_beam}), we can write (\ref{eq:optimal_gamma_ext}) more intuitively as
\begin{equation*}
\gamma^\star_\ext = \gamma_0 + \gamma_\text{beam}.
\end{equation*}
\thesisnote{We will make further use of definition (\ref{eq:def_gamma_beam}) in Section~\ref{sec:gamma_ext_dependence}.}
\end{remark}

\begin{remark}
There are two different conventions for the synchronous phase $\phibo$, i.e., the nominal phase between the particle bunches and the accelerating mode \cite{Tuckmantel2011}. For linacs and circular electron machines it is conventionally defined so that 
$\phi_\text{b0,lin} = \pi - \angle \Ibo$ \cite{Schilcher1998,Padamsee2008,Wangler2008}, which gives the well-known expressions
\begin{alignat*}{2}
\Dw^\star &= - \frac{\alpha I_{b0} \sin\phibolin}{2A_0}, \quad &&
\gamma_\ext^\star = \gamma_0 + \frac{\alpha I_{b0} \cos \phibolin}{2A_0}. \\
\intertext{For circular proton machines \cite{Tuckmantel2011,Wiedemann2007}, the convention is such that $\phi_\text{b0,circ} = 3\pi/2 - \angle \Ibo$, which gives}
\Dw^\star &= \frac{\alpha I_{b0} \cos\phibocirc}{2A_0}, \quad && 
\gamma_\ext^\star = \gamma_0 + \frac{\alpha I_{b0} \sin \phibocirc}{2A_0}.
\end{alignat*}

\end{remark}

\begin{remark}
	For pulsed linacs, the value of $\gamma_\ext$ that minimizes the overall power consumption is somewhat larger than $\gamma_\ext^\star$ since this gives a shorter filling time.
	This is particularly important when the pulses are short compared to the filling time.
	It is also better to choose $\gamma_\ext$ larger than $\gamma_\ext^\star$ if there are significant detuning variations during the flat-top (e.g., from microphonics) \cite{Merminga1996}.
\end{remark}

\begin{remark}
The coupling factor $\beta = Q_0 / Q_\ext = \gamma_\ext / \gamma_0$ is commonly used in previous derivations of optimal coupling \cite{Wangler2008,Padamsee2008,Merminga1996,Schilcher2007}. For perfectly superconducting cavities ($\beta = \infty$), many expressions in those derivations are ill-defined. The derivation in this section avoids that unaesthetic feature.
\end{remark}

\begin{remark}
The detuning angle $\psi = \tan^{-1}(\Dw / \gamma)$ is often used for describing the steady-state response of the accelerating mode. Note that for the transfer function $P_a(s)$ in \eqref{eq:tf_acc_mode} we have $\Pa(0) = \cos \psi \cdot \me^{i\psi}$.
\end{remark}

\section{Normalized cavity dynamics}
\label{sec:normalized_cav_model}

\subsection{Normalization}
Requirements on cavity-field errors, and specifications on amplifier ripple and beam-current ripple are typically given in relative terms, i.e., on the form $x \si{\percent}$ and $y \si{\degree}$.
By normalizing the dynamics from disturbances to field errors (of the accelerating mode), it is easy to compute the relative field errors that result from relative disturbances.

For control design, it is convenient if the static gain from control action to mode amplitude is one.

For these reasons, we introduce the following normalized phasors for the cavity field, \rf{} drive, and beam loading, colored according to \cref{fig:phasor_plot_example},
\begin{subequations}
	\begin{align}
	{\color{field-color} \ba} &\coloneqq \,\, \frac{1}{A_0} \, {\color{field-color} \bA} \\[0.2em]
	{\color{ig-color} \fg} &\coloneqq \frac{1}{\gamma A_0} {\color{ig-color}  \sqrt{2\gamma_\ext}\Fg } \\[0.2em]
	{\color{ib-color} \ib} &\coloneqq \frac{1}{\gamma A_0} {\color{ib-color}  \frac{\alpha\Ib}{2} }.
	\end{align}
	\label{eq:cavity_dynamics_normaliztion}%
\end{subequations}
Scaling equation (\ref{eq:cav_eq_proposed}) by $1/A_0$ gives
\begin{equation}
\boxed{
	\dot{\color{field-color}\ba} =
	(-\gamma + i\Dw) \ba
	+ \gamma({\color{ig-color} \fg} + {\color{ib-color}  \ib}).
}
\label{eq:normalized_cav_eq}
\end{equation}

The transfer function from $\fg$ and $\ib$ to $\ba$ is given by
\begin{equation}
\Pa(s) \coloneqq \frac{\gamma}{s + \gamma - i\Dw},
\label{eq:tf_acc_mode}
\end{equation}
where the subscript $a$ indicates the accelerating mode.

\begin{remark}
The relative beam-loading parameter $Y$ in \cite{Pedersen1975} corresponds to $\abs{\ibo}$.
\end{remark}

\subsection{Relations at nominal operating point}
Consider steady-state operation at some nominal operating point $(\ba_0\!=\!1, \fgo, \ibo)$.
For an \emph{optimally tuned} cavity it follows from (\ref{eq:optimal_detuning}) that
\begin{equation}
\gamma\Im\, {\color{ib-color}\ibo} + \Dw = 0.
\end{equation}
For an \emph{optimally coupled} cavity, it follows from (\ref{eq:optimal_gamma_ext}) that
\begin{equation}
-1 \leq \Re\, {\color{ib-color} \ibo} \leq 0.
\end{equation}
For an \emph{optimally tuned and optimally coupled} cavity we have that $\fgo = 1 - i\Dw/\gamma - \ibo$ is real, and that
\begin{equation}
1 \leq {\color{ig-color} \fgo} \leq 2.
\end{equation}

For a superconducting cavity ($\gamma_0 = 0$) that is optimally tuned and coupled, we have that $\Re\, \ib = -1$ and $\fg = 2$.

\subsection{Dynamics around nominal operating point}
Cavity field stability is typically evaluated around some nominal operating point.
For this reason it is meaningful to introduce the normalized field error $\bz$ through
\[
\bz = \ba - 1.
\]
It is clear that a small field error \mbox{$\bz= z_\re + iz_\im$} approximately corresponds to an amplitude error of $z_\re \cdot 100 \,\si{\percent}$ and a phase error of $z_\im \, \si{\radian}$.

In the case of an ideal amplifier, we would have $\fg = \fgo + \fgtilde$ where $\fgtilde$ corresponds to control action from the field controller.
However, due to variations of the amplifier's gain and phase shift, denoted by $\tilde{g}_\text{amp}$ and $\tilde{\theta}_\text{amp}$, we have
\begin{equation}
\fg = (1 + \tilde{g}_\text{amp})\me^{-i\tilde{\theta}_\text{amp}} (\fgo + \fgtilde).
\label{eq:fg_expansion1}
\end{equation}
Assuming that the variations $\tilde{g}_\text{amp}$ and $\tilde{\theta}_\text{amp}$ are small, and introducing
$\bdg \coloneqq \tilde{g}_\text{amp} - i\tilde{\theta}_\text{amp}$, it follows from \eqref{eq:fg_expansion1} that
\begin{equation}
\fg
\approx (1 + \bdg) (\fgo + \fgtilde) \approx \fgo + \fgtilde + \fgo\bdg.
\label{eq:fg_expansion2}
\end{equation}
Similarly, relative beam loading variations $\bdb$ affect $\ib$ according to
\[
\ib = (1 + \bdb)\ibo.
\]

Plugging these expressions into (\ref{eq:normalized_cav_eq}) and ignoring second-order terms give
\begin{equation}
\dot{\bz} =
(-\gamma + i\Dw) \bz
+ \gamma \left( \fgtilde + {\color{ig-color} \fgo \bdg}
+ {\color{ib-color} \ibo \bdb}
\right).
\label{eq:normalized_cav_eq_linearized}
\end{equation}

We see that the transfer function from relative disturbances $\bdg$ and $\bdb$ to relative field errors $\bz$ are given by%
\begin{subequations}\label{eq:tf_dist_to_err}%
	\begin{align}
	{\color{ig-color} P_{d_\text{g} \rightarrow z}}(s) &= {\color{ig-color} \fgo} \Pa(s), \label{eq:tf_amp_dist_to_err} \\
	{\color{ib-color} P_{d_\text{b} \rightarrow z}}(s) &= {\color{ib-color} \ibo} \Pa(s),
	\label{eq:tf_beam_dist_to_err}
	\end{align}
\end{subequations}
where $\Pa(s)$ is defined in \eqref{eq:tf_acc_mode}.
In these equations, the nominal phasors $\fgo$ and $\ibo$ act as complex-valued coefficients that quantify the impact of relative disturbances.

\medskip
\noindent\emph{Example:} The normalized cavity parameters $\gamma$, $\fgo$, and $\ibo$ for the cavities in Table~\ref{tab:cav_params_physical} are shown in Table~\ref{tab:cav_params_normalized}.
\begin{table}
	\centering
	\setlength{\tabcolsep}{5pt}
	\caption{Normalized parameters for the cavities in Table~\ref{tab:cav_params_physical}.}
	
	\begin{tabular}{l@{\hskip 20pt}cccc}
		\toprule
		& $\gamma/2\pi$  & $\fgo$ & $\abs{\ibo}$ & $\phi_{b0}$ \\
		\textbf{Cavity} & \si{\kilo\hertz} & $-$ & $-$ & \si{\degree} \\
		\midrule
		\ESSRFQ & 60  & 1.1 & 0.2 & $-45$ \\
		\ESSDTL & 12  & 1.3 & 0.3 & $-25$ \\
		\ESSMedBeta & 0.5  & 2.1 & 1.1 & $-15$ \\
		\XFELTesla & 0.14  & 2.0 & 1.0 & $\approx\!0$ \\
		\LCLSIITesla & 0.016 & 1.3 & 0.3 & $\approx\!0$ \\
		\midrule		
		\CESR & 1.2 & 2.0 & 3.0 & 70 \\
		\bottomrule
	\end{tabular}
	\label{tab:cav_params_normalized}
\end{table}

\begin{remark} 
In superconducting cavities, the detuning $\Dw$ may vary due to microphonics and Lorenz-force detuning.
If we denote the detuning variations by $\widetilde{\Dw}$ we can define the normalized detuning variations $\dDw \coloneqq {\widetilde{\Dw}}/\gamma$.
If the detuning variations are small, they can be included as a term $i\gamma \cdot \dDw$ on the right-hand side of \eqref{eq:normalized_cav_eq_linearized}. The transfer function from normalized detuning variations to field errors is given by	$P_{\dDw \rightarrow z}(s) = i\Pa(s)$.

\end{remark}

\begin{remark} 
The disturbances $\bdg$ and $\bdb$ tend to have a certain directionality. For example, \rf{} amplifiers such as klystrons, typically have more phase variations than amplitude variations, corresponding to that $\bdg$ is dominantly imaginary.
Beam-current ripple on the other hand affects the magnitude of $\ib$ (corresponding to a real $\bdb$).
The directionality of these disturbances are readily visualized as in \cref{fig:phasor_plot_directionality}, using the phasor diagrams of Section~\ref{sec:phasor_diagram}.
Synchrotron oscillations in circular accelerators correspond to an imaginary $\bdb$.

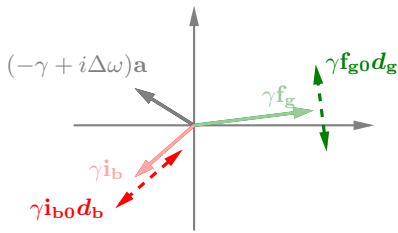
\begin{figure}
	\centering
	\begin{tikzpicture}
\newcommand{\ibcoord}{-0.8, -0.7}%
\newcommand{\decaycoord}{-0.8, 0.5}%

\newcommand{\captionDerivativePhasorPlot}{}
\drawderivativediagram

\draw[phasor,black!50] (O2) -- ++(decay) node[above left, xshift=3mm] {$(-\gamma + i\Dw)\ba$};
\draw[phasor,ib-color!35] (O2) -- ++(ib) node[left, yshift=1mm] {$\gamma\ib$};
\draw[phasor, ig-color!40]     (O2) -- ++(ig) node[above left, xshift=-1mm, yshift=-1mm] {$\gamma\fg$};

\draw[<->, very thick, dashed, ib-color] ($(O2)+0.3*(ib) + (0.1, -0.1)$) -- ($(O2) + 1.45*(ib) + (0.1, -0.1)$) node[left] {$\gamma \ibo \bdb$};

\coordinate (ig-absolute) at ($(O2)+(ig)$);
\draw[<->, very thick, dashed, ig-color] ($(ig-absolute)!0.6cm!90:(O2)  + (0.09, 0.03)$) -- ($(ig-absolute)!0.6cm!-90:(O2) +  (0.09, 0.03)$) node[right] {$\gamma \fgo \bdg$};

\end{tikzpicture}
	\caption{Visualization of phase variations of the \rf{} drive ($\bdg$ purely imaginary) and amplitude variations of the beam current ($\bdb$ purely real).
	The effect on the accelerating cavity mode is given by filtering the illustrated variations through the transfer function $\Pa(s)$ in \eqref{eq:tf_acc_mode}.\thesisnote{S(s)?}}
	\label{fig:phasor_plot_directionality}
\end{figure}

\end{remark}

\section{Summary}
We have proposed an energy-based parameterization of a cavity's accelerating-mode dynamics.
The proposed parameterization avoids many problems of equivalent-circuit based parameterizations and is helpful for understanding the impact of cavity parameters on the \rf{} system and field control loop.
We have also provided a normalized model that is suitable for field control design.

\begin{acknowledgments}
The author thanks Daniel Sjöberg,  Larry Doolittle, and Bo Bernhardsson for helpful comments and suggestions. The author is a member of the ELLIIT Strategic Research Area at Lund University.	
\end{acknowledgments}

\section*{Appendix: Derivation of Equation (\ref{eq:cav_eq_proposed})}
\appendix 

We show how equation \eqref{eq:cav_eq_proposed} follow from conservation of energy together with linearity and time-reversibility of Maxwell's equations. The considered system, together with the notation that will be used, are shown in \cref{fig:cavity_waveguide}.

Note that the derivation in this appendix is not intended to prove a new result; equation \eqref{eq:cav_eq_proposed} simply follows from \eqref{eq:cav_eq_tuckmantel}/\eqref{eq:cav_eq_schilcher} by using  Table~\ref{tab:cavity_parameter_comparison}. The derivation is instead included for an easy-to-follow connection between the model \eqref{eq:cav_eq_proposed} and the physical cavity--waveguide system.

We start by going through Haus' derivation for the dynamics of a waveguide-coupled cavity \cite[Sec. 7.2]{Haus1983}, using the notation of this paper.
Then we show how the impact of beam loading can be included.
It is assumed that: (1) changes of the mode shape due to resistive losses, the external coupling, and detuning variation can be neglected; (2) the slowly-varying envelope approximation holds.
For modeling of multiport cavities, see \cite{Suh2004}.

\subsection{Maxwell's equation for the electromagnetic field}
We start by considering a lossless cavity without beam connected to a waveguide, and later introduce losses and beam loading in sections \ref{sec:waveguide_coupling} and \ref{sec:beam_loading}.
From Maxwell's equations, we have that the following equation holds for the electric field  $\bEtv = \bEtv(\br, t)$ in the cavity and the waveguide
\begin{equation}
\nabla^2 \bEtv  - \epsilon_0\mu_0 \pdd{}{t} \bEtv= 0,
\label{eq:em_wave_equation}
\end{equation}
where $\varepsilon_0$ and $\mu_0$ are the permittivity and permeability of free space.

\subsection{Mode expansion of the cavity field}
Assume for a moment that the cavity is not coupled to the waveguide. The electric field in the cavity can then be expanded as a sum of orthogonal eigenmodes $\bE_k$
\begin{equation}
\bEtv(\br, t) =  \sum_{k=0}^\infty e_k(t) \bE_k (\br)
\label{eq:mode_expansion}
\end{equation}
where the mode amplitudes $e_k(t)$ evolve independently according to
\begin{equation}
\dd{}{t} e_k(t) = -\omega_k^2 e_k(t).
\label{eq:mode_evolution_rf}
\end{equation}

\subsection{Baseband dynamics of the accelerating mode\\in a lossless cavity}

We will only consider the specific mode used for particle acceleration.
When necessary, we will label related quantities with a subscript $a$.
To simplify the exposition, and keep with the spirit of the paper, we will work with the complex envelope $\bA$ of the accelerating mode (relative to some phase reference with frequency $\omega_\RF$); i.e., $e_a(t) = \Re \{ \bA(t) \me^{i\omega_\RF t}  \}$.
We will also assume that the mode amplitude is normalized so that $\abs{\bA}^2$ equals the energy stored in the mode ($\bA$ has units $\sqrt{\si{J}}$).
From (\ref{eq:mode_evolution_rf}) it follows that
\begin{equation}
\frac{d}{dt} \bA = i\Dw \bA
\label{eq:cav_eq_isolated}
\end{equation}
where $\Dw \coloneqq \omega_a - \omega_\RF$.

\subsection{Waveguide coupling}
\newcommand{\kappag}{\bm{\kappa}_\mathrm{g}}
\label{sec:waveguide_coupling}
Now, assume that the cavity is connected to a waveguide by a coupling port as in  \cref{fig:cavity_waveguide}.
An incident forward wave in the waveguide will excite the accelerating mode through the coupling port, but energy will also escape the cavity through the port and propagate away in a reverse wave (\cref{fig:cavity_waveguide}).
Denote the complex-envelopes (with respect to $\omega_\RF$) of the forward and reverse waves by $\Fg$ and $\Rg$, and assume them normalized so that $\abs{\Fg}^2$ is the power of the forward wave ($\Fg$ has units $\sqrt{\si{W}}$) and similarly for $\Rg$.

Due to the linearity of Maxwell's equations, we have
\begin{equation}
\frac{d}{dt} \bA = i\Dw\bA -\gamma_\ext \bA + \kappag \Fg
\label{eq:cav_eq_undetermined}
\end{equation}
where $\gamma_\ext$ is the rate at which the cavity field decays through the coupling port, and $\kappag$ is a, possibly complex-valued, parameter that quantifies the effect of the forward wave on the cavity field.
Not surprisingly, $\gamma_\ext$ and $\kappag$ are related and we next derive how.

\subsection{Relation between $\bm{\gamma}_\ext$ and $\bm{\kappa}_\textbf{g}$}
\label{sec:relation_gamma_ext_kappa}
Throughout this subsection we consider the particular solution to \eqref{eq:cav_eq_undetermined} for $t \geq 0$ that is given by
the initial condition $\bA(0) = 1$, with $\Dw = 0$ and $\Fg(t) \equiv 0$.
It is clear that the solution is given by
\begin{equation}
\bA(t) = \me^{-\gamma_\ext t}, \qquad t\geq 0.
\label{eq:cav_derivation_specific_sol}
\end{equation}

Recall that the energy stored in the accelerating mode is $\abs{\bA}^2$ and hence changes by $\frac{d}{dt} \abs{\bA}^2  = -2\gamma_\ext \me^{-2\gamma_\ext t}$. Due to conservation of energy, this power is carried away by the reverse wave, hence
\begin{equation}
\abs{\Rg(t)}^2 = 2\gamma_\ext \me^{-2\gamma_\ext t}.
\label{eq:energy_reverse_wave}
\end{equation}

If $\bEtv(\br, t)$ is a solution to \eqref{eq:em_wave_equation}, valid in the cavity and the waveguide, then so is the time-reversed solution $\bEtv_r(\br, t) = \bEtv(\br, -t)$.

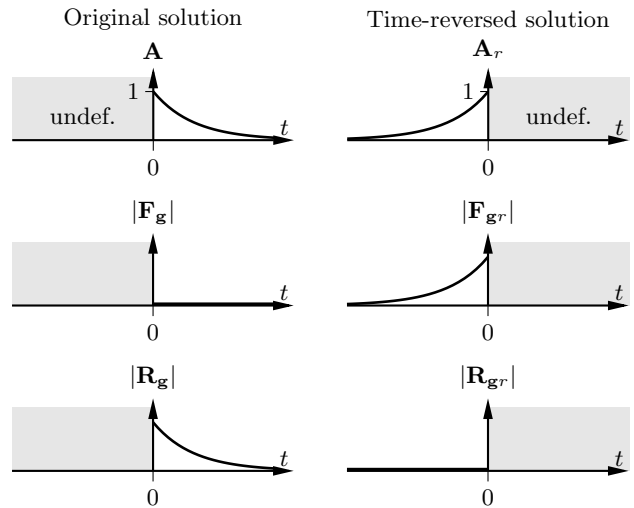
\begin{figure}
	\centering
	\newcommand{\FillStart}{0.27}%
\pgfmathsetmacro{\FTstart}{\FillStart+0.69}%
\newcommand{\FTend}{2.0}%
\newcommand{\Xmax}{3.5}%
\newcommand{\Ymaxshade}{1.3}
\begin{tikzpicture}[clip=false]
\begin{groupplot}[%
group style={group size=2 by 3, vertical sep=35pt, horizontal sep=20pt},
arrowplotLongLabel,  
width=3.75cm,
height=0.97cm,
scale only axis,
xmin=-\Xmax,
xmax=\Xmax,
ymin = 0,
ymax = 1.5,
xtick={0},
ytick={0},
xlabel={$t$},
xlabel style={at={(1.2,0.1)}},
xlabel style={xshift=-25pt,yshift=8},
ylabel style={yshift=10pt},
title style={yshift=-6pt},
axis y line=middle,
axis on top,
axis line style={thick,->},
ytick pos=left,
ytick style={draw=none} 
]

\nextgroupplot[title={$\bA$}, ytick={1}, clip=false]	
\fill[gray!20] (axis cs: 0,0) rectangle (axis cs: -\Xmax,\Ymaxshade);	
\addplot[domain=0:\Xmax-0.5] {exp(-\x)} node[above]{};
\addplot[thin,domain=-0.2:0] {1};

\node at (axis cs: 0, 2.5) {Original solution};

\node[align=center] at (axis cs: -\Xmax/2, 0.5) {undef.};

\nextgroupplot[title={$\bArev$}, ytick={1}, clip=false]
\fill[gray!20] (axis cs: 0,0) rectangle (axis cs: \Xmax,\Ymaxshade);	
\addplot[domain=-\Xmax:0] {exp(\x)} node[above]{};
\addplot[thin, domain=-0.2:0] {1};

\node at (axis cs: 0, 2.5) {Time-reversed solution};

\node[align=center] at (axis cs: \Xmax/2, 0.5) {undef.};

\nextgroupplot[title={$|\Fg|$}]
\fill[gray!20] (axis cs: 0,0) rectangle (axis cs: -\Xmax,\Ymaxshade);	
\addplot[ultra thick, domain=0:\Xmax-0.5] {0.02} node[above]{};

\nextgroupplot[title={$|\Fgrev|$}]
\fill[gray!20] (axis cs: 0,0) rectangle (axis cs: \Xmax,\Ymaxshade);	
\addplot[domain=-\Xmax:0] {exp(\x)} node[above]{};

\nextgroupplot[title={$|\Rg|$}]
\fill[gray!20] (axis cs: 0,0) rectangle (axis cs: -\Xmax,\Ymaxshade);	
\addplot[domain=0:\Xmax-0.5] {exp(-\x)} node[above]{};
\coordinate (normal-equation) at (axis cs: 0,-1.7);

\nextgroupplot[title={$|\Rgrev|$}]
\fill[gray!20] (axis cs: 0,0) rectangle (axis cs: \Xmax,\Ymaxshade);	
\addplot[ultra thick, domain=-\Xmax:0] {0.02} node[above]{};
\coordinate (reversed-equation) at (axis cs: 0,-1.7);

\end{groupplot}

\end{tikzpicture}
	\caption{The particular solution considered in Sec.~\!\ref{sec:relation_gamma_ext_kappa}.}
	\label{fig:time_reversed_solution}
\end{figure}

Time-reversal of the particular solution considered in this subsection gives a solution where the evolution of the accelerating mode is given by
\begin{equation}
\bArev(t) = \bA(-t) =  \me^{\gamma_\ext t},  \quad t \leq 0,
\label{eq:reversed_mode_amplitude}
\end{equation}
the reverse wave satisfies $\Rgrev(t) = \Fg(t) \equiv 0$, and the forward wave satisfies $|\Fgrev(t)| = \abs{\Rg(-t)}$. From \eqref{eq:energy_reverse_wave} we then have that
\begin{equation}
\Fgrev(t) =  \me^{i\phi_0} \cdot \sqrt{2\gamma_\ext} \me^{\gamma_\ext t}
\label{eq:reversed_fwd_wave}
\end{equation}
for some phase $\phi_0$. The variables of the original solution and time-reversed solution are illustrated in \cref{fig:time_reversed_solution}.

Recalling that $\Dw=0$, and plugging \eqref{eq:reversed_mode_amplitude} and \eqref{eq:reversed_fwd_wave} into \eqref{eq:cav_eq_undetermined} gives
\[
\gamma_\ext  \me^{\gamma_\ext t} = -\gamma_\ext  \me^{\gamma_\ext t}
+ \kappag \me^{i\phi_0} \cdot \sqrt{2\gamma_\ext} \me^{\gamma_\ext t},
\]
from which it follows that
\[
\kappag =  \me^{-i\phi_0} \cdot \sqrt{2\gamma_\ext}.
\]

The reference phase for the forward wave $\Fg$ can be chosen freely; choosing it such that $\phi_0 = 0$ gives
\[
\kappag = \sqrt{2\gamma_\ext}.
\]

\subsection{Beam loading}
\label{sec:beam_loading}

Consider a beam, i.e., a train of charged bunches, traversing the cavity.  Assume that the bunches are regularly spaced in time by an integral number of \rf{} periods.
Let the beam be modeled by the complex signal $\Ib$ whose magnitude $\abs{\Ib}$ equals the \dc{} current of the beam.
The phase of $\Ib$ is defined  so that $\angle \Ib = -\pi$  corresponds to maximum acceleration (energy gain) from the nominal field of the accelerating mode (corresponding to $\angle \bA = 0$). Note that $\Ib$ is allowed to vary slowly.

Define the cavity--beam-coupling parameter $\alpha$ of the accelerating mode so that the following equality holds (in \cite[Ch. 2]{Wangler2008} it is shown that such an $\alpha$ exists)
\begin{equation}
\parbox{4.3cm}{power to beam from accelerating mode} = - \Re\{ \alpha \Ib^* \bA \}.
\label{eq:power_to_beam}
\end{equation}

The cavity--beam-coupling parameter of the accelerating mode is real and non-negative due to the definition of the $\Ib$. For a general mode $k$, the cavity--beam-coupling parameter is in general complex and the $\alpha$ on the right-hand side of \eqref{eq:power_to_beam} should be replaced by  $\bm{\alpha}_k^*$.

\newcommand{\bcb}{\bm{c}_{\mathbf{b}}}

The bunch train induces an electromagnetic field in the cavity, corresponding to a term $-\mu_0 \,d \bJ / dt$, where $\bJ$ is current density, on the right-hand side of \eqref{eq:em_wave_equation}. This effect is linear and corresponds to a term $\bcb \Ib$, where $\bcb$ is a complex coefficient, on the right-hand side of \eqref{eq:cav_eq_isolated}.
Assuming for a moment that $\Dw = 0$, we have
\[
\frac{d\bA}{dt} = \bcb \Ib.
\]
Taking the time derivative of the energy in the accelerating mode and using this expression we get that
\begin{equation}
\frac{d}{dt} \abs{\bA}^2 = 2  \Re \{ \bcb^* \Ib^* \bA \}.
\label{eq:energy_decreae_beam_loading}
\end{equation}

Conservation of energy gives that \eqref{eq:power_to_beam} and \eqref{eq:energy_decreae_beam_loading} sum to zero. Since this holds for all $\Ib$ it follows that $\bcb = \alpha/2$.

\subsection{Putting the pieces together}
By combing the results from the two preceding sections and adding a term $-\gamma_0 \bA$ for resistive losses (assuming that this does not significantly change the mode shape) we arrive at
\begin{equation}
	\d{\bA}{t}  = (-\gamma + i\Dw) \bA  + \sqrt{2\gamma_\ext} \Fg + \frac{\alpha}{2}  \Ib
\label{eq:baseband_cav_eq_final}
\end{equation}
where $\gamma = \gamma_0 + \gamma_\ext$. This is exactly \eqref{eq:cav_eq_proposed}.

\subsection{The reverse wave}\label{sec:reverse_wave}

As in \cite[(7.36)]{Haus1983} we may derive an expression for the envelope $\Rg$ of the reverse wave.
From the linearity of Maxwell's equations we know that the reverse wave depends linearly on the forward wave and the cavity field
\[ \Rg = \bm{c}_F \Fg + \bm{c}_A \bA, \]
where $\bm{c}_F$ and $\bm{c}_A$ are complex constants.
We already know from \eqref{eq:cav_derivation_specific_sol} and  \eqref{eq:energy_reverse_wave} that $\abs{\bm{c}_A} = \sqrt{2\gamma_\ext}$. Since we are free to choose the reference phase for $\Rg$ we will take $\bm{c}_A = \sqrt{2\gamma_\ext}$.

Next, we derive an expression for $\bm{c}_F$.
Conservation of energy gives that
\begin{equation}
\abs{\Fg}^2 - \abs{\Rg}^2 = \d{}{t} \abs{\bA}^2,
\label{eq:bwdwave_deriv1}
\end{equation}
and from (\ref{eq:baseband_cav_eq_final}) (with $\Ib = 0$) it follows that
\begin{equation}
\d{}{t} \abs{\bA}^2 = -2\gamma_\ext \abs{\bA}^2 + \sqrt{2\gamma_\ext}\Big( \bA^* \Fg + \Fg^*\bA  \Big).
\vspace*{0.5em}
\label{eq:bwdwave_deriv2}
\end{equation}
Putting (\ref{eq:bwdwave_deriv1}) equal to (\ref{eq:bwdwave_deriv2}), and then substituting $\bA = (\Rg -\bm{c}_F \Fg)/\sqrt{2\gamma_\ext}$  gives
\begin{multline*}
\abs{\Fg}^2 - \abs{\Rg}^2 = -2\gamma_\ext \abs{\bA}^2 + \sqrt{2\gamma_\ext}\Big( \bA^* \Fg + \Fg^* \bA \Big)
\\
= -  \left(\abs{\Rg}^2
-\bm{c}_F \Rg^*\Fg - \bm{c}_F^* \Fg^* \Rg
+ \abs{\bm{c}_F}^2\abs{\Fg}^2 \right) \\
 \hspace*{1cm} + \Big( \Rg^*\Fg -\bm{c}_F^* \abs{\Fg}^2
+ \Fg^*\Rg -\bm{c}_F \abs{\Fg}^2  \Big).
\end{multline*}
From this equation it follows that
\begin{equation*}
\abs{ (1 + \bm{c}_F) \Fg - \Rg}^2 = \abs{\Rg}^2.
\end{equation*}
For this equality to hold for all $\Fg$ and $\Rg$, we must have that $\bm{c}_F = -1$ and hence the reverse wave is given, as in \cite[(7.36)]{Haus1983}, by
\[
\Rg = -\Fg + \sqrt{2\gamma_\ext} \bA.
\]

\bibliography{acc_mode_modeling_refs}
\onecolumngrid 

\end{document}